\documentclass[a4paper,11pt]{article}
\usepackage{pos}
\usepackage{subfigure}
\usepackage{slashed}

\title{Background processes in Higgs decay to Z gamma}

\author*[a]{Aliaksei Kachanovich}

\affiliation[a]{Service de Physique Théorique, CP225,\\
Université Libre de Bruxelles \\
Boulevard du Triomphe (Campus de la Plaine) \\
1050 Bruxelles
Belgium}

\emailAdd{aliaksei.kachanovich@ulb.be}

\abstract{The ATLAS and CMS Collaborations reported that the observed number of Higgs boson decays into a $Z$ boson and a photon is $\mu = 2.2 \pm 0.7$ times higher than predicted by the Standard Model. Initially, this discrepancy was attributed to a modification of the $HZ\gamma$ vertex. In the $H \to Z\gamma$ process, this decay is reconstructed from $H \to \ell\ell\gamma$, where $\ell$ represents either an electron or a muon. In this study, an investigation is conducted to examine this anomaly by exploring potential additional background contributions to $H \to \ell\ell\gamma$ from various subprocesses within and beyond the Standard Model.}

\FullConference{
9th Symposium on Prospects \\ in the Physics of Discrete Symmetries \\
December 2-6, 2024 \\
Ljubljana
}

\begin{document}
\maketitle

\section{Introduction}
The Standard Model (SM) is extremely successful but not a complete theory. Significant efforts are dedicated to searching for discrepancies with the SM. After the discovery of the Brout-Englert-Higgs boson \cite{ATLAS:2012yve,CMS:2012qbp} (commonly referred to as the Higgs boson), many of its properties remain unknown. Recent results from the ATLAS and CMS collaborations provide the first evidence for the rare decay  $H \rightarrow Z\gamma$ \cite{ATLAS:2020qcv,CMS:2022ahq}. The combined measurements from both experiments yield a branching fraction of  $\text{B}_{\rm obs} = (3.4 \pm 1.1) \times 10^{-3} $, which exceeds the Standard Model prediction by a factor of $( 2.2 \pm 0.7 )$ \cite{ATLAS:2023yqk}. The discrepancy corresponds to only a $1.9\sigma$ standard deviation. This observation has motivated numerous studies aimed at providing an explanation (e.g. \cite{Barducci:2023zml,Boto:2023bpg,Das:2024tfe}).

Experimentally, the process $ H \to Z \gamma $ is reconstructed by measuring the $\ell \ell \gamma$ final state, where $\ell = e, \mu$. The background for $H \to Z \gamma \to \ell \ell \gamma$ includes contributions from $H \to \gamma \gamma$, box diagrams, and the tree-level decay of the Higgs boson into a muon pair, $H \to \mu \mu \gamma$ \cite{Kachanovich:2020xyg,Kachanovich:2021pvx,Corbett:2021iob,Chen:2021ibm,Ahmed:2023vyl,Hue:2023tdz,VanOn:2021myp,Sun:2013cba,Phan:2021xwc,Phan:2021ovj,Kachanovich:2024vpt,Abbasabadi:1996ze,Chen:2012ju,Dicus:2013ycd,Passarino:2013nka,Han:2017yhy}.  The reconstruction of $H \to Z \gamma$ involves applying kinematic cuts on the dilepton invariant mass above $50$ GeV \cite{ATLAS:2020qcv,CMS:2022ahq,ATLAS:2023yqk}, which significantly reduces the contribution of the $H \to \gamma \gamma$ background but does not eliminate it completely (see Fig.~\ref{fig:Rescale}). The currently reported $H \to Z \gamma$ signal strength is $\mu = 2.2 \pm 0.7$ times the SM prediction.

The new approach to resolving this excess involves introducing a new background provided by BSM physics processes in Higgs decay \cite{Kachanovich:2025cxz}.

\section{The Standard Model}
The final state $\ell \ell \gamma$ is produced by 4-different sub-processes: tree level decay with Bremsstrahlung, one-loop $H \to  \gamma \gamma \to \ell \ell \gamma$, direct box-diagram coupling, and one-loop $H \to  Z\gamma \to \ell \ell \gamma$. One can write the total one-loop amplitude as
\begin{eqnarray}
    \mathcal{M}_{\text{SM,loop}} &=& \nonumber \left[q_{\mu} p_{1} \cdot \varepsilon^{*}(q) - \varepsilon^{*}_\mu(q) \, q\cdot p_{1}\right] \bar{u}(p_{2}) \big( a_{1} \gamma^{\mu} P_{R} + b_{1} \gamma^{\mu} P_{L} \big) v(p_{1}) \\
    &+& \left[q_{\mu} p_{2} \cdot \varepsilon^{*}(q) - \varepsilon^{*}_\mu(q) \, q\cdot p_{2}\right] \bar{u}(p_{2})\big( a_{2} \gamma^{\mu} P_{R} + b_{2} \gamma^{\mu} P_{L} \big) v(p_{1}) \,,
    \label{eq:SMloop}
\end{eqnarray}
The four-momenta of the photon and leptons are denoted by $q$, $p_{1}$, and $p_{2}$, respectively.

The form factors $a_{1(2)}$ and $b_{1(2)}$ depend on the Mandelstam variables, defined as $s = (p_1 + p_2)^2$, $t = (p_1 + q)^2$, and $u = (p_2 + q)^2$, which satisfy the relation $s + t + u = m_H^2 + 2 m_{\ell}^2 \approx m_H^2$, where $m_H$ is the $H$ boson mass and $m_\ell$ is the lepton mass. The coefficients $a_{1(2)}$ are symmetric under the interchange of $t$ and $u$, as are the coefficients $b_{1(2)}$. Explicit expressions for the SM coefficients $a_{1,2}$ and $b_{1,2}$ can be found in \cite{Kachanovich:2020xyg}.

Resonant and non-resonant contributions are defined by separating the form factors in Eq.~\ref{eq:SMloop} into two subsets (for more details, see \cite{Kachanovich:2021pvx}):  

\begin{center}
\begin{figure}[t]
\centering
{\includegraphics[width=0.7\textwidth]{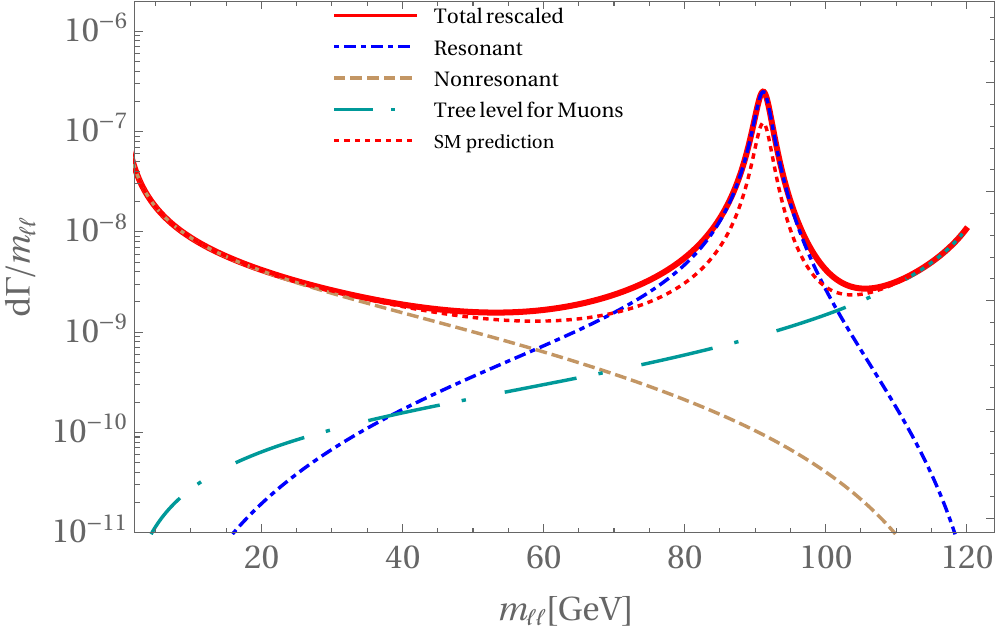}}
	\caption{The total contribution with the modified $HZ\gamma$ vertex is shown by the solid red curve, while the blue dash-dotted line represents the rescaled resonant contribution of the SM. This is compared to the SM prediction (short-dashed red curve). The brown dashed line indicates the non-resonant contribution, and the long dash-dotted green line represents the tree-level contribution.} 
	\label{fig:Rescale}
\end{figure}
\end{center}

\begin{eqnarray}
    a_{1(2)} (s,t) =  a_{1(2)}^{res}(s) + a_{1(2)}^{nr}(s,t) \,, 
\end{eqnarray}
with  
\begin{eqnarray}
     a_{1(2)}^{res}(s) &\equiv& \frac{\alpha (m_{Z}^2)}{s - m_{Z}^2 + i m_{Z} \Gamma_Z}, \qquad \mbox{\rm and} \qquad  
     a_{1(2)}^{nr}(s,t) \equiv \tilde{a}_{1(2)} (s, t) + \frac{\alpha(s) - \alpha (m_{Z}^2)}{s - m_{Z}^2 + i m_{Z} \Gamma_Z}\,,
\end{eqnarray}  
where $\alpha (s)$ is a component of the form factors $a_{1(2)}$, which remains the same in both coefficients. A similar decomposition holds for the form factors $b_{1(2)}$. The original form factor can then be rewritten as  
\begin{eqnarray}
    a_{1(2)}(s,t) = \tilde{a}_{1(2)}(s,t) + \frac{\alpha (s)}{s - m_{Z}^2 + i m_{Z} \Gamma_Z}\,.
\end{eqnarray}  
with a similar expression for $b_{1(2)}(s,t)$.
The resonant contribution depends solely on the dilepton squared invariant mass, denoted by the variable $s$.  

The excess in $H \to \ell \ell \gamma$ can be explained by a modification of the $H \to Z \gamma$ vertex, that represented by rescaling of the resonant part (see Fig.~\ref{fig:Rescale}).

\section{New Physics}
\subsection{Effective Field Theory}

\begin{figure}[]
	\begin{center}
		\subfigure[t][]{\includegraphics[width=0.49\textwidth]{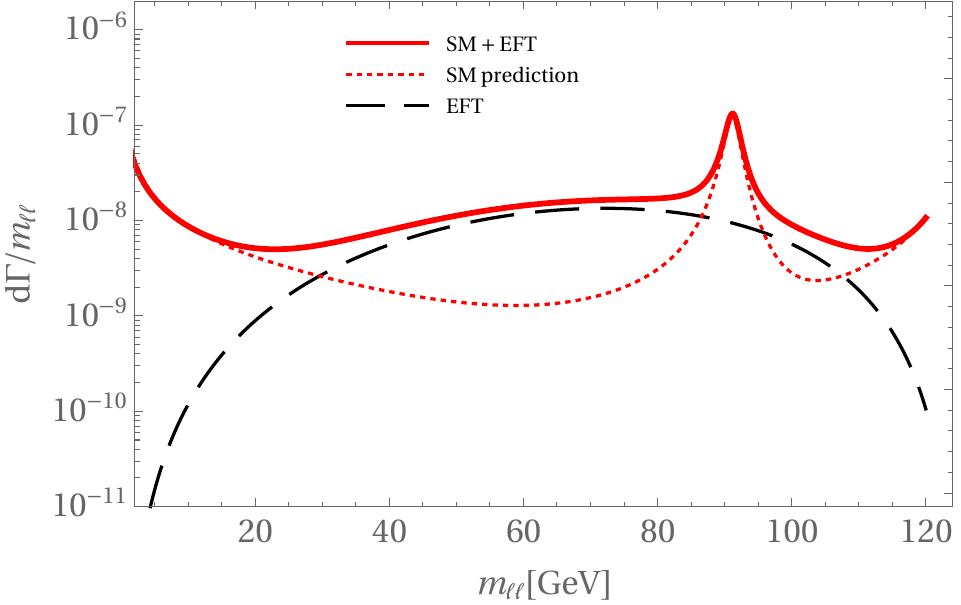}}
		\hspace{.1cm}
  	\subfigure[t][]{\includegraphics[width=0.49\textwidth]{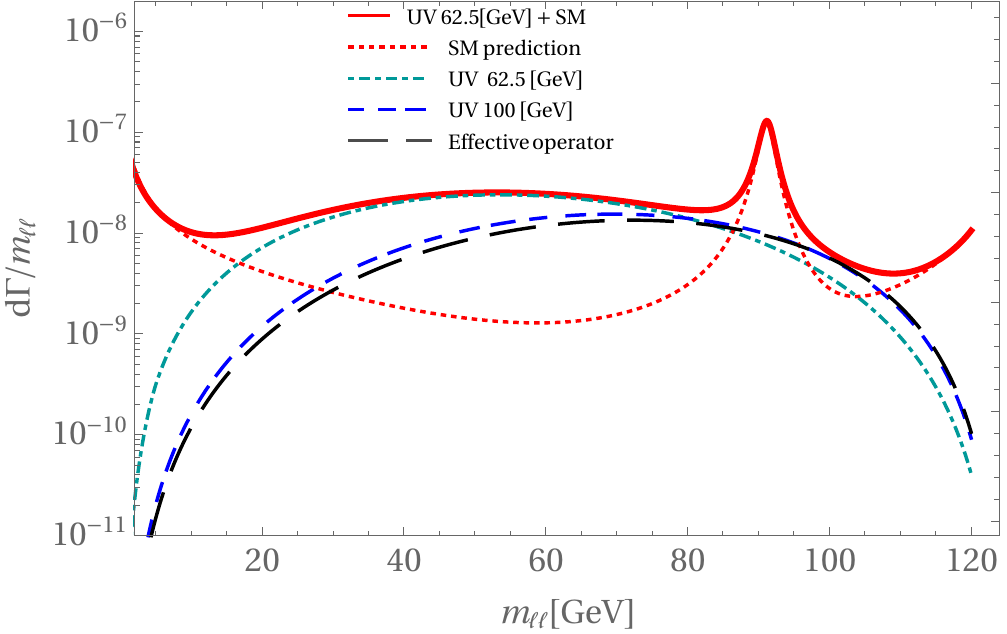}}
         \end{center}
	\caption{Panel (a) shows the contribution of the effective operator Eq.~\ref{eq:Lag_EFT} to $H \rightarrow \ell \ell \gamma$ is considered. The new physics scale is set to $\Lambda_R = 260$ GeV to account for the observed excess of events compared to the SM prediction. Panel (b) shows possible UV-complete contributions to the differential decay rate for two benchmark candidates are shown: UV $6.25$ (dot-dashed green and solid red) with masses $M = m_S = m_{\Psi} = 62.5$ GeV, and UV $100$ GeV with $M = 100$ GeV. The black long-dashed curve corresponds to the effective operator, as in panel (a).}
	\label{fig:BSM_tot}
\end{figure}
One of the possible solutions for the additionally observed events, without modifying the $HZ\gamma$ vertex, is the introduction of an additional background. The most general way to incorporate BSM physics contributions is through the introduction of an Effective Field Theory (EFT) operator. One such operator can be a dimension-8 operator

\begin{equation}\label{eq:Lag_EFT}
   {\cal L}_{\rm eff} \supset {g'\over \Lambda_R^4} \vert \Phi  \vert^2 \partial_{\nu} (\bar \ell_{R} \gamma_{\mu} \ell_R) B^{\mu \nu} \,,
\end{equation}

\noindent where $\Phi$ is the SM Higgs doublet, $\ell_R$ are $SU(2)$ singlet right-handed leptonic spinors, $B^{\mu\nu}$ is the hypercharge field strength, and $g' = e/\cos\theta_W$. The dimension-6 part on which this operator is built, $\partial_{\nu} (\bar{\ell}_R \gamma_{\mu} \ell_R) B^{\mu \nu}$, vanishes for on-shell fields; therefore, the factor $\vert \Phi \vert^2$ plays a crucial role.

\subsection{UV-complete model}

As a possible UV-complete model, scenarios with additional scalar and lepton fields have been considered \cite{Toma:2013bka,Giacchino:2013bta}
\begin{equation}
    \mathcal{L}_{UV} \supset \frac{1}{2} \partial_\mu S \partial^\mu S - \frac{1}{2} m_S^{2} S^{2} + \bar{\Psi} (i \slashed{D} - m_{\Psi})\Psi - \sum_{\ell} (y_{\ell} S \bar{\Psi} \ell_{R} + h.c.) - \frac{\lambda_{hs}}{2} S^2 |H|^{2} 
    \label{eq:LagUV}
\end{equation}

\noindent where $S$ is a real scalar, singlet under the SM. Potentially, $S$ can serve as a DM candidate \cite{Silveira:1985rk, McDonald:1993ex, Burgess:2000yq}. It interacts with the SM through the Higgs portal and also couples to right-handed leptons via Yukawa interactions with a vector-like fermion $\Psi$, which carries hypercharge $Y = Q = -1$.  

\begin{figure}[ht]
	\begin{center}
		\subfigure[t][]{\includegraphics[width=0.23\textwidth]{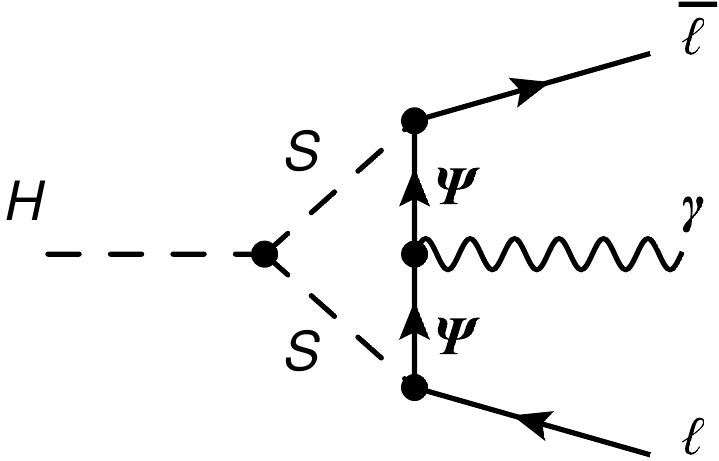}}
		\hspace{.6cm}
  	\subfigure[t][]{\includegraphics[width=0.23\textwidth]{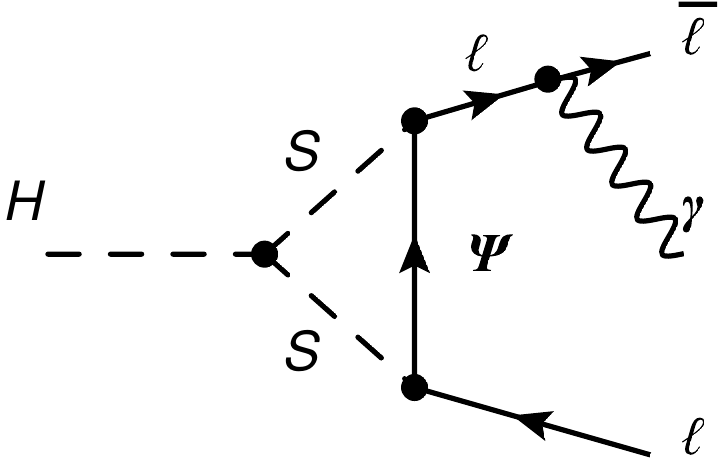}}
   		\hspace{.6cm}
   		\subfigure[t][]{\includegraphics[width=0.23\textwidth]{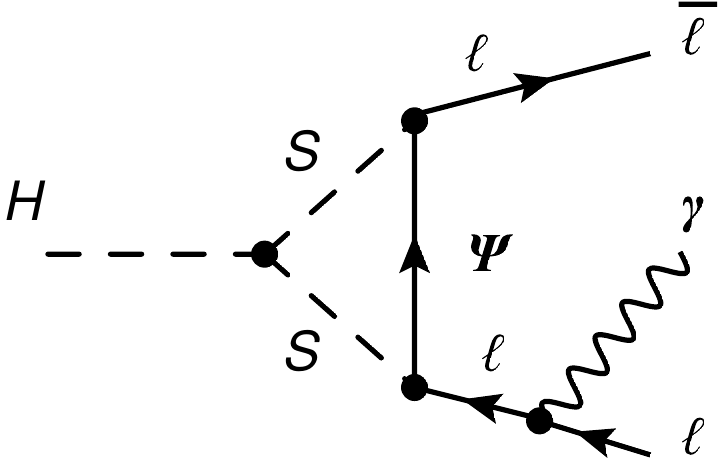}}
         \end{center}
	\caption{Amplitudes for the UV model: Bremsstrahlung from final-state leptons (b, c) is loop- and chirality-suppressed, making its contribution negligible for electrons and muons. In the chiral limit, the virtual fermion emission (a) matches the structure of the effective operator Eq.~\ref{eq:Lag_EFT}. }
	\label{fig:Loop_diagram}
\end{figure}

With this Lagrangian, the extra contribution to the SM in the process $H \to \ell \ell \gamma$ is represented by three additional diagrams (Fig. \ref{fig:Loop_diagram}), which were evaluated using \emph{FeynCalc} \cite{Shtabovenko:2023idz}.

\section{Phenomenology}

The background required to achieve the observed decay rate $\Gamma_{\rm obs} = 0.93$ keV for the process $H \to \ell \ell \gamma$ is obtained for the EFT operator (Eq.~\ref{eq:Lag_EFT}) at the scale $\Lambda = 260$ GeV. 

The UV-complete theory has four free parameters: two coupling constants ($y_{\ell}$ and $\lambda_{hs}$) and two masses ($m_{\Psi}$ and $m_{S}$). The required background can be obtained with different values of these parameters. Two benchmark scenarios are considered: one with both masses set to  $m_{\Psi} = m_{S} = 62.5$ GeV (half of the Higgs mass) and another with both at  $m_{\Psi} = m_{S} = 100$ GeV. In both cases, the Higgs-scalar coupling is fixed at $\lambda_{hs} = 0.26$, corresponding to couplings of $y_{\ell} = 1.66$ and $y_{\ell} = 10.50$, respectively.

Experimental kinematic cuts play a crucial role. All scenarios reproduce the number of events for the CMS cuts ($E_\gamma \geq 15$ GeV, $E_1 \geq 7$ GeV, $E_2 \geq 25$ GeV, and $t_{\rm min}, u_{\rm min} \geq (0.1 m_H)^2$). Tab.~\ref{tab:sample} presents the predicted decay rates for different cut choices across various scenarios.

\begin{table}[ht!]
\centering
\begin{tabular}{|c|c|c|c|c|c|c|c|c|}
\hline
\,\#\,& {Cuts} & $m_{\ell \ell}^{min}$[GeV] & $m_{\ell \ell}^{max}$[GeV]  & $\Gamma_{\rm tot}^{\rm SM}$ [keV] & $\Gamma^{\rm SM}_{\rm tree}$[keV] & {$\frac{Br_{\rm resc}}{Br_{\rm SM}}$} & {$\frac{Br_{\rm EFT}}{Br_{\rm SM}}$} & {$\frac{Br_{\rm UV}}{Br_{\rm SM}}$}\\
\hline
1 & \, None \, & 50 & 125  & 0.768 & 0.287 & 1.67 & 1.86 & 2.07\\
\hline
2 & None & 50 & 100  & 0.504 & 0.028 & 2.01 & 2.21 & 2.57\\
\hline
3 & CMS & 40 & 125  & 0.455 & 0.011 & 2.04 & 2.10 & 2.13 \\
\hline\hline
\textbf{4} & \textbf{CMS} & \textbf{50} & \textbf{125} & \textbf{0.451} & \textbf{0.011}  & \textbf{2.06} & \textbf{2.06} & \textbf{2.06} \\
\hline\hline
5 & CMS & 70 & 125   & 0.440 & 0.011  & 2.07 & 1.80 & 1.71\\
\hline
6 & CMS & 70 & 100   & 0.432 & 0.006  & 2.08 & 1.74 & 1.68 \\
\hline
7 & CMS & 80 & 100   & 0.416 & 0.005  & 2.09 & 1.48 & 1.39\\
\hline
\end{tabular}
\caption{Signal strengths for various cut choices are presented. The first two lines assume no cuts on the kinematic parameters, except for $m_{\rm \ell\ell}$, where the value was obtained by integrating over the $t$-Mandelstam variable. The CMS cuts are taken as: $E_\gamma \geq 15$ GeV, $E_1 \geq 7$ GeV, $E_2 \geq 25$ GeV, and $t_{\rm min}, u_{\rm min} \geq (0.1 m_H)^2$. The column $Br_{\rm resc}/Br_{\rm SM}$ corresponds to the expected signal strength from the $Z$ peak over that of the SM, assuming the effective $H Z \gamma$ coupling is rescaled by a factor of $\sqrt{2.11}$ to obtain the observed branching fraction $\text{B}{\rm obs} = (3.4 \pm 1.1) \times 10^{-3}$. The column $Br_{\rm EFT}/Br_{\rm SM}$ corresponds to the signal strength using the effective operator (Eq.\ref{eq:Lag_EFT}). The column $Br_{\rm UV}/Br_{\rm SM}$ corresponds to the UV model (Eq.\ref{eq:LagUV}) with $m_S = m_\Psi = 62.5$ GeV.}
\label{tab:sample}
\end{table}

\section{Conclusions}

This work explores the possibility that new physics could contribute to the background in measurements of the $HZ\gamma$ effective coupling, motivated by indications of an excess reported by both the ATLAS and CMS collaborations. BSM physics is considered both in terms of an effective operator (Eq.~\ref{eq:Lag_EFT}) and a simplified UV model (Eq.~\ref{eq:LagUV}) with a scale close to the electroweak scale, which is also motivated by the dark matter problem.

The differential decay rate for the rescaled $HZ\gamma$ vertex Fig.~\ref{fig:Rescale}, is compared to the EFT and UV-complete model background contribution Fig.~\ref{fig:BSM_tot} (all plots are shown without kinematic cuts). The signal signatures differ significantly in the proposed models, requiring further experimental investigation across dilepton mass bins $m_{\ell \ell}$. Various scenarios with kinematic cuts on $m_{\ell \ell}$, including experimental selections, are summarized in Tab.~\ref{tab:sample}. 

It is deemed unlikely that new physics significantly contributes to $H \rightarrow \ell^+\ell^- \gamma$. The UV model considered is ad hoc and fine-tuned, relying on several unnatural assumptions, including a low new physics scale, a compressed mass spectrum ($m_S \approx m_F$), and identical Yukawa couplings for electrons and muons. Nevertheless, other experimental constraints are considered, along with the possibility that one of the new particles could contribute to dark matter. The model should remain relevant for future analyses, even if the excess disappears with the same cuts.

\section*{Acknowledgements}
I thank Jean Kimus, Steven Lowette, and Michel H.G. Tytgat for the enjoyable collaboration on this work. I also thank Giorgio Arcadi, Debtosh Chowdhury, and Laura Lopez Honorez for helpful discussions, as well as Ivan Nišandžić for providing the code used to cross-check parts of our work. This research was supported by FRS/FNRS, FRIA, the BLU-ULB Brussels Laboratory of the Universe, and the IISN convention No. 4.4503.15.

\bibliographystyle{apsrev4-1}

\bibliography{main}



\end{document}